# IMPROVEMENT OF ANOMALY DETECTION ALGORITHMS IN HYPERSPECTRAL IMAGES USING DISCRETE WAVELET TRANSFORM

Mohsen Zare Baghbidi[1], Kamal Jamshidi[1], Ahmad Reza Naghsh Nilchi[1] and Saeid Homayouni[2]

[1]Department of Computer Engineering, College of Engineering, Isfahan University, Isfahan, Iran
`{m_zare, jamshidi, nilchi}@eng.ui.ac.ir`
[2] Remote Sensing Group, Department of Geomatics, College of Engineering, University of Tehran
`homayounis@ut.ac.ir`

## ABSTRACT

*Recently anomaly detection (AD) has become an important application for target detection in hyperspectral remotely sensed images. In many applications, in addition to high accuracy of detection we need a fast and reliable algorithm as well. This paper presents a novel method to improve the performance of current AD algorithms. The proposed method first calculates Discrete Wavelet Transform (DWT) of every pixel vector of image using Daubechies4 wavelet. Then, AD algorithm performs on four bands of "Wavelet transform" matrix which are the approximation of main image. In this research some benchmark AD algorithms including Local RX, DWRX and DWEST have been implemented on Airborne Visible/Infrared Imaging Spectrometer (AVIRIS) hyperspectral datasets. Experimental results demonstrate significant improvement of runtime in proposed method. In addition, this method improves the accuracy of AD algorithms because of DWT's power in extracting approximation coefficients of signal, which contain the main behaviour of signal, and abandon the redundant information in hyperspectral image data.*

## 1. INTRODUCTION

Recently hyperspectral imaging has been recognized as a suitable tool for target detection and recognition in many applications including search-and-rescue operations, mine detection and military usages. Hyperspectral sensors collect valuable information of earth's surfaces in hundreds of narrow contiguous spectral bands in the visible and infrared regions of the electromagnetic spectrum. These information providing a powerful means to discriminate different materials on the basis of their unique spectral signatures [1].

Anomaly detection (AD) is a particular case of target detection (TD) with no *a priori* information about targets. The main goal of AD algorithms is finding the objects that are anomalous with respect to the background [1]. To be more precise, the aim of AD algorithms is to find the pixels whose spectra differ significantly from the background spectra [2]. The power of anomaly detection technique is that we do not need to target signature and atmospheric/radiometric corrections [3].

      

In recent years, many hyperspectral AD algorithms have been proposed including Reed-Xialoi (RX) family algorithms, like DWRX and Kernel-RX, and other algorithms like DWEST and NSWTD [1], [4]. RX is considered as a benchmark AD algorithm for hyperspectral images [5]. The most well known problem for RX family algorithms is the small sample size. This problem comes from the estimation of a local background covariance matrix from a small number of very high dimensional samples. As a result, the covariance matrix of local background will be bad-conditioned and its estimation is unstable which strongly affects the detection performance of AD algorithm [6]. In addition, in many applications (like real time applications) runtime of the algorithm is an important issue which has not been considered by the proposed AD algorithms. An efficient solution for these problems is applying the dimensionality reduction (DR) techniques [7].

DR can be done as a pre-processing step for AD algorithms. This method reduce interband spectral redundancy and improve the separation between anomaly and background signatures, so improve detection performance of anomaly detector [8]. There are two categories of DR techniques: linear and nonlinear. Linear techniques do not exploit nonlinear properties in hyperspectral image, but they can be fast enough for real time application. Some linear DR techniques like Principle Component Analysis (PCA) are very popular and widely used in hyperspectral applications [9]. A newer DR method is discrete wavelet transform (DWT). This method did not evaluate for improvement of algorithms in the AD literature.

This paper presents a new DR method using DWT which acts as a pre-processing step for AD algorithm. This method overcomes expressed drawbacks and improves the performance and runtime of anomaly detectors.

The paper is organized as follows. Section 2 provides an overview of three popular AD methods used in this study. DWT is introduced in section 3. Section 4 presents a brief description of the proposed method. Experimental results will discuss in section 5. Finally, Section 6 concludes the paper.

## 2. ANOMALY DETECTION ALGORITHMS

### 2.1. RX Detector (RXD)

RX is a widely used anomaly detector which was developed by Reed and Yu in [5]. It's a constant false alarm rate (CFAR) detector and derived from the generalized likelihood ratio test. It also considered as a benchmark AD algorithm for hyperspectral images and works as follows:

Assume that *r* is an image pixel vector that has *L* elements (*L* is image's spectral bands). RX anomaly detector defines by equation (1). Where *μ* is the sample mean and *C* is the sample data covariance matrix.

$$\delta_{rxd}(r) = (r - \mu)^T C_{L \times L}^{-1} (r - \mu) \tag{1}$$

AD algorithms like RX can be grouped in two categories: global and local. Global anomaly detectors define *background* with reference to all the image pixels. In local case *background* is defined in a small neighborhood of pixel under test. Covariance matrix is calculated according to the defined *background*. In this study RX is implemented locally and named Local RX (LRX).

### 2.2. DWEST algorithm

Dual window-based eigen separation transform (DWEST) is an adaptive anomaly detector which was developed by Kwon et al [10]. It implements two windows, called inner and outer windows which are designed to maximize the separation between two-class data. These two classes are target and background classes. DWEST can briefly describe as follows:

Assume that $r$ is an image pixel at which inner and outer windows are centered. Let $m_{outer}(r)$ and $m_{inner}(r)$ is the means of the outer and inner windows respectively and $C_{outer}$ and $C_{inner}$ is their respective covariance matrices. $C_{diff}$ is the difference covariance matrix between $C_{outer}$ and $C_{inner}$ and defined by equation (2). As a result, the eigen values of $C_{diff}$ can be divided into two groups: negative and positive values. Kwon et al.'s argued that the eigenvectors associated with a small number of the large positive eigen values of $C_{diff}$ could successfully extract the spectrally distinctive materials that are present in the inner window. If the eigenvectors represented by the positive eigen values in this small set are denoted by $\{v_i\}$, the anomaly detector derived by the DWEST ($\delta_{DWEST}(r)$) projects the differential means of two windows ($m_{diff}(r)$, which showed in equation (3)) onto $\{v_i\}$ specified by equation (4).

$$C_{diff} = C_{inner} - C_{outer} \tag{2}$$

$$m_{diff} = m_{inner} - m_{outer} \tag{3}$$

$$\delta^{DWEST}(r) = \left| \sum_{v_i} v_i^T m_{diff}(r) \right| \tag{4}$$

## 2.3. DWRX algorithm

This algorithm is the combination of RX and DWEST. In other words, it's a RX detector which is implemented by two windows, inner and outer window that defined in part 2.2. This detector is defined by equation (5). Parameters here are the same as DWEST definition [11].

$$\delta^{|RXD|}(r) = \left| m_{diff}(r)^T [C_{outer}^{-1}(r)] m_{diff}(r) \right| \tag{5}$$

## 3. DISCRETE WAVELET TRANSFORM

The foundation of the discrete wavelet transform (DWT) goes back to 1976 when Crochiere et al. for the first time introduced sub-band coding [12]. In 1983, Burt defined a technique very similar to sub-band coding and named it pyramidal coding which is also known as multi-resolution analysis [13]. Later in 1989, Vetterli and Le Gall made some improvements to the sub-band coding scheme and removed the existing redundancy in the pyramidal coding scheme [14]. DWT definition is based on sub-band coding and multi-resolution analysis.

In DWT the procedure starts with passing the main signal through a half-band digital low-pass filter with impulse response $h[n]$. The output of filter is convolution of the signal with impulse response of the filter as shown in equation (6):

$$x[n] * h[n] = \sum_{k=-\infty}^{\infty} x[k].h[n-k] \tag{6}$$

Passing the signal ($x[n]$) through a half-band low-pass filter removes all frequencies that are above half of the highest frequency in the main signal. According to the Nyquist's rule, half of the samples can be eliminated. This procedure is done by down-sampling (or sub-sampling) of the output of low-pass filter by two as shown in equation (7).

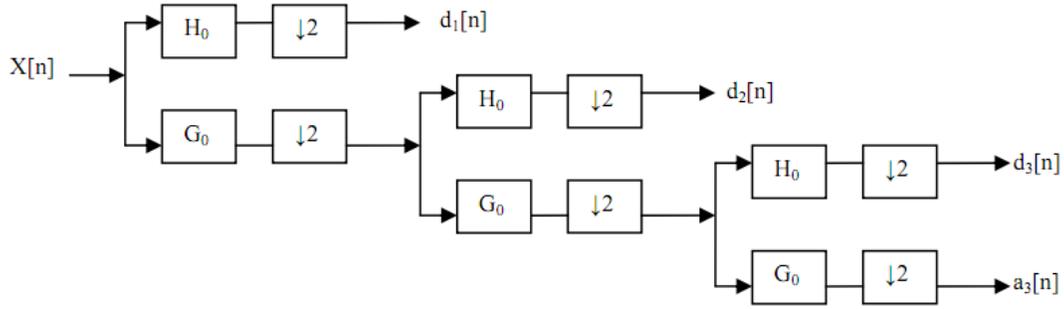

Figure 1. Three-level wavelet decomposition tree

$$y[n] = \sum_{k=-\infty}^{\infty} h[k].x[2n-k] \tag{7}$$

The same procedure is done by a half-band high-pass filter with impulse response *g[n]*. This procedure constitutes one level of decomposition in DWT and can be expressed mathematically by equations (8) and (9) where $y_{high}[k]$ and $y_{low}[k]$ are the output of high-pass and low-pass filters after down-sampling [15].

$$y_{high}[k] = \sum_{n} x[n].g[2k-n] \tag{8}$$

$$y_{low}[k] = \sum_{n} x[n].h[2k-n] \tag{9}$$

This decomposition, which is known as sub-band coding, halves the time resolution and doubles the frequency resolution and can be repeated for further decomposition by filtering the output of low-pass filter as shown in Figure 1 for three level of decomposition [16].

Levels of decomposition for DWT are related to filters and frequency specifications of main signal. Finally the DWT of original signal is obtained by concatenation of all coefficients, starting from the last level of decomposition. So the DWT coefficients are equal to original signal's coefficients. According to Figure 1, output of DWT is [$a_3.d_3.d_2.d_1$].

## 4. PROPOSED METHOD

When the DWT coefficients of a signal is calculated, the frequencies that are the most prominent in the original signal will be appear as high amplitude in related regions of the DWT signal. Frequencies that are not prominent will have very low amplitude in DWT signal and can be discarded without any major loss of information.

A pixel of a hyperspectral image is a vector with *L* elements, which *L* is number of image's spectral bands. Low frequencies in this signal are the most prominent and high frequencies can be related to noise. So the main behaviour of signal can be found in approximation coefficients of DWT (output of the low-pass filter).

Figure 2 shows a 64-sample signal which is the spectrum for one pixel in a hyperspectral image with 64 spectral bands and its DWT's coefficients which has been calculated by different types of Daubechies wavelet [17]. Figure 2.b shows the 2-level DWT of the signal which calculated by Daubechies16 wavelet. The last 32 samples in this signal (Figure 2.b) correspond to the highest frequency bands in the signal; the previous 16 samples correspond to the second highest frequency bands and of the signal and the first 16 samples correspond to low frequencies of original signal which are approximation coefficient of it. Figure 2.c and Figure 2.d shows the 3-level and 4-level DWT of the signal respectively.

Four-level DWT (Figure 2.d) is calculated using Daubechies4 wavelet. Only the first 4 samples in this signal, which correspond to low frequencies of the original signal, carry relevant information and the rest of this signal has virtually no information. Therefore, all but the first 4 samples can be discarded without any loss of information. These 4 samples are the approximation coefficient of main signal. In the proposed method these 4 samples, is used to detect anomalies.

The proposed method first calculates DWT of every pixel of hyperspectral image using Daubechies4 wavelet. Daubechies4 wavelet decomposes the signal until four samples are left. These four samples are calculated for every pixel and placed in a matrix named "approximation matrix". Approximation matrix is abstract of the main image and has the main behaviour of it. Then AD algorithms are performed on this matrix instead of main hyperspectral image.

## 5. EXPERIMENTAL RESULTS

### 5.1. Hyperspectral data

To evaluate the performance of proposed method, two datasets is used: image with implanted targets and image with real targets. These data are extracted from a Hyperspectral image of a naval air station in San Diego, California, collected by AVIRIS (Airborne Visible and InfraRed Imaging Spectrometer) sensor. This image has 189 useful spectral bands and its ground resolution is 3.5 meters.

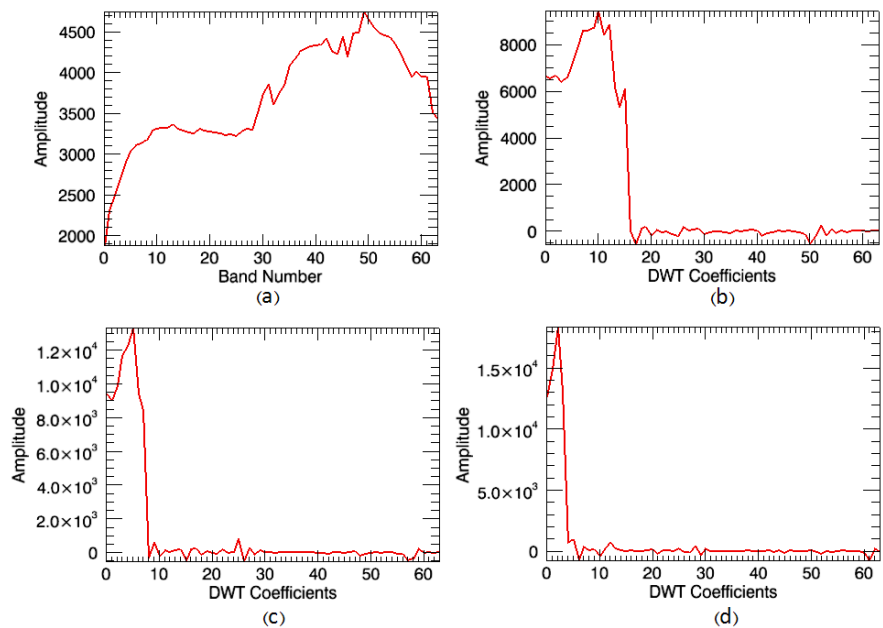

Figure 2. (a) Original signal, (b) 2-level DWT, (c) 3-level DWT, (d) 4-level DWT

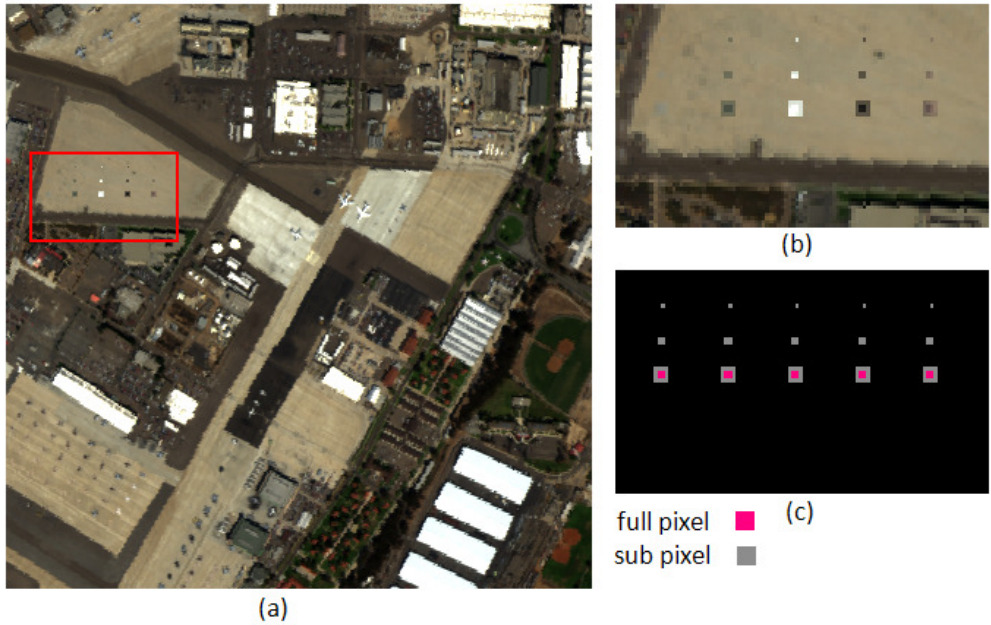

Figure 3. (a) A natural colour composite of the AVIRIS image, (b) subimage with implanted targets (Img1), and (c) truth locations of targets in Img1.

### 5.1.1. Image with Implanted targets

To accurately evaluate the performance of anomaly detectors, an image with truth location of targets (anomalies) is needed. For this purpose a 60×100 pixels sub-image (from main AVIRIS image, Figure 3.a) has been selected and the target implanted method [18–20] has been used to implant targets in this sub-image (named Img1, see Figure 3.c).

In target implanted method, a sub-pixel synthetic anomaly, *z*, is a combination of target and background as shown in equation (10). In this equation *t* and *b* denotes target and background respectively. Sub-pixel (*z*) consists of the target's spectrum, with fraction *f,* and the background's spectrum, with fraction (*1-f*) [19].

$$z = f.t + (1-f).b \qquad (10)$$

This method does not include adjacency effects of target spectrum on the local background pixels. To be more real, the background pixels which are neighbours of the targets can be affected by target pixel. This effect can be done by a Gaussian function with width of *w* as shown in equation (11), where $p_i$ is spatial distance between the target pixel (*t*) and background pixel ($z_i$) [21]:

$$z_i = \exp\left(-\frac{\rho_i^2}{w^2}\right).f.t + \left(1 - \exp\left(-\frac{\rho_i^2}{w^2}\right).f\right).b_i \qquad (11)$$

To implant anomalies in the sub-image, different targets from various parts of main image has been selected and implanted in the sub-image as shown in Figure 3.b. To apply the effect of background on targets, outlines of targets has been selected and combined with their adjacent background according to equation (10) with coefficient *f=0.55*. To apply the effect of anomalies on the background pixels (at a spatial distance of one from the target) equation (11) is used. This data cube includes both subpixel and multi-pixel targets, so it's an excellent image for testing AD algorithms. The truth location of targets is shown in Figure 3.c.

#### 5.1.2. Image with real targets

To evaluate the runtime and performance of algorithms on a hyperspectral image with real targets, a sub-image with size of 100×100 was selected from AVIRIS data cube. In this sub-image there are 38 anomalous targets. These targets may be helicopters or helipads (Figure 4.b). This sub-image which is used in works like [22] and [23], named Img2.

### 5.2 Implementation

To evaluate the proposed method three anomaly detection algorithms, Local RX (LRX), DW-RX and DWEST, in standard mode and with proposed method have been implemented by IDL programming language. To address AD algorithms which have been implemented with proposed pre-processing method, they named DWT-LRX, DWT-DWRX and DWT-DWEST.

An important decision for AD algorithms is the detection windows size. There is no specific method to choose these windows [21], but inner window size (in dual window algorithms) should be almost as large as the biggest target in the image. The size of outer window should be large enough to have sufficient number of background sample for further processing [24].

To implement LRX and DWT-RX algorithms a window of 15×15 pixels has been used for both Img1 and Img2. The size of inner and outer window for DWRX, DWT-DWRX, DWEST and DWT-DWEST algorithms have been selected 5×5 and 13×13 pixels for Img1 and 3×3 and 13×13 pixels for Img2, respectively.

### 5.3 Evaluation of algorithms

There are two important specifications for evaluation of anomaly detectors: accuracy of detection and runtime. In this study the accurate performance of AD algorithms is evaluated using Img1 and Img2 is used for performance evaluation virtually and runtime investigation of anomaly detectors.

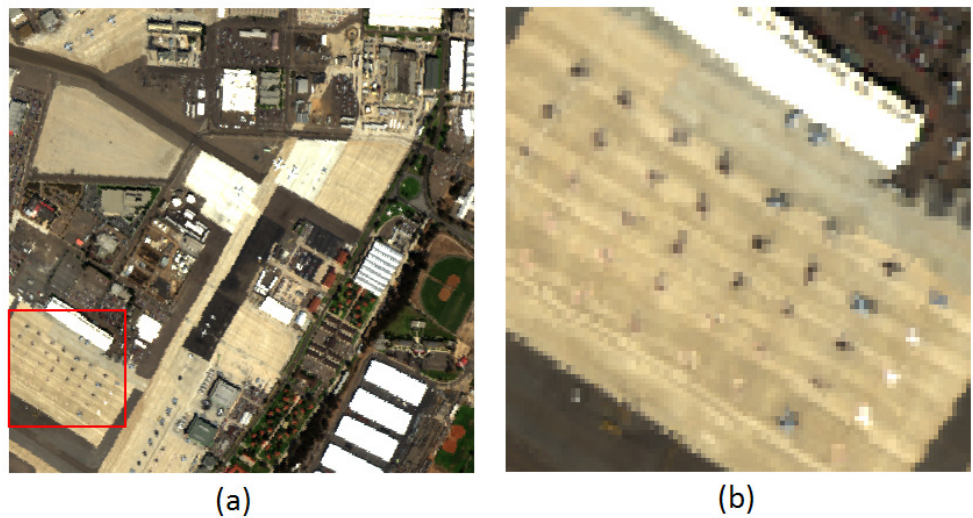

Figure 4. A natural colour composite of the AVIRIS image, (b) sub-image with real targets (Img2)

### 5.3.1. Performance evaluation by Img1

The accuracy of AD algorithms is evaluated based on the receiver operation characteristic (ROC) curve. ROC curve shows true detection (TD) rate versus false alarm rate (FAR) over a particular scenario. TD and FAR are computed by varying the detection threshold and counting the number of true detected targets and the corresponding number of false alarms [1]. The area under ROC curves (AUC) is an exact criterion and widely accepted to evaluate and compare detection exactness of AD methods [7].

Figure 5 shows the detection results of AD methods on Img1. ROC curves of AD algorithms are showed in Figure 6. AUC values for AD methods are showed in Table 1 and Figure 7 compare anomaly detectors performance according to their AUCs. According to these results, detection performance of LRX and DWRX algorithms is significantly improved using proposed pre processing method and the performance of DWEST method is not changed. In addition, DWT-DWRX is the best detector.

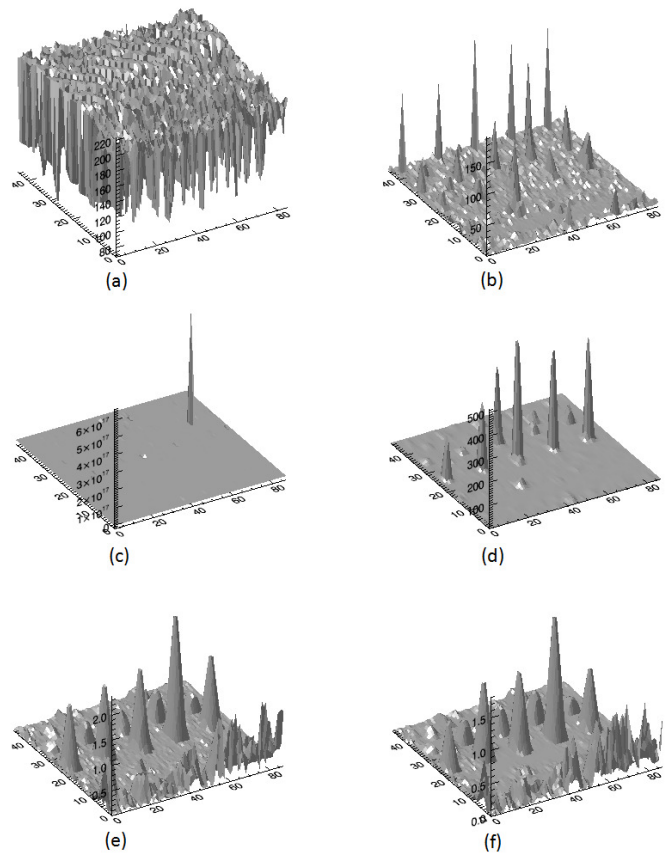

Figure 5. 3D plots of detection results for Img1: (a) LRX, (b) DWT-LRX, (c) DWRX, (d) DWT-DWRX, (e) DWEST and (f) DWT-DWEST

Signal & Image Processing : An International Journal (SIPIJ) Vol.2, No.4, December 2011

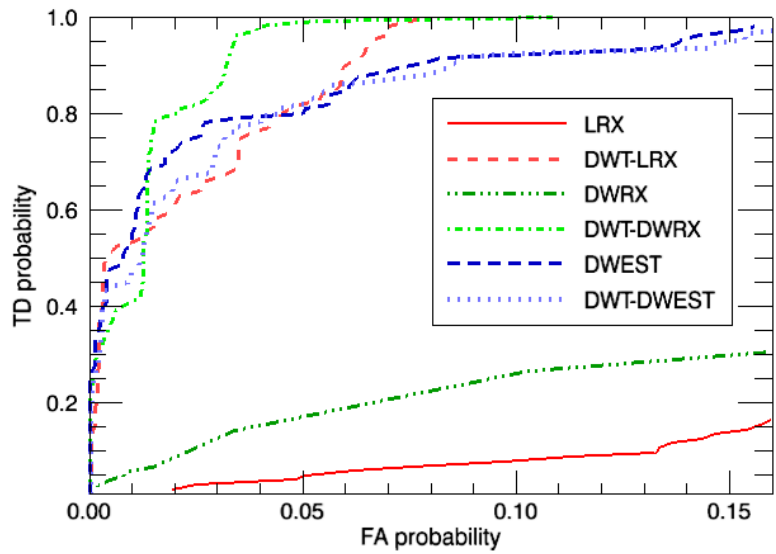

Figure 6. ROC curves of AD methods for Img1.

Table 1. AUCs of AD methods for Img1.

| AD algorithm | AUC |
|---|---|
| *LRX* | 0.5412 |
| *DWT-LRX* | 0.9791 |
| *DWRX* | 0.5271 |
| *DWT-DWRX* | **0.9872** |
| *DWEST* | 0.9745 |
| *DWT-DWEST* | 0.9709 |

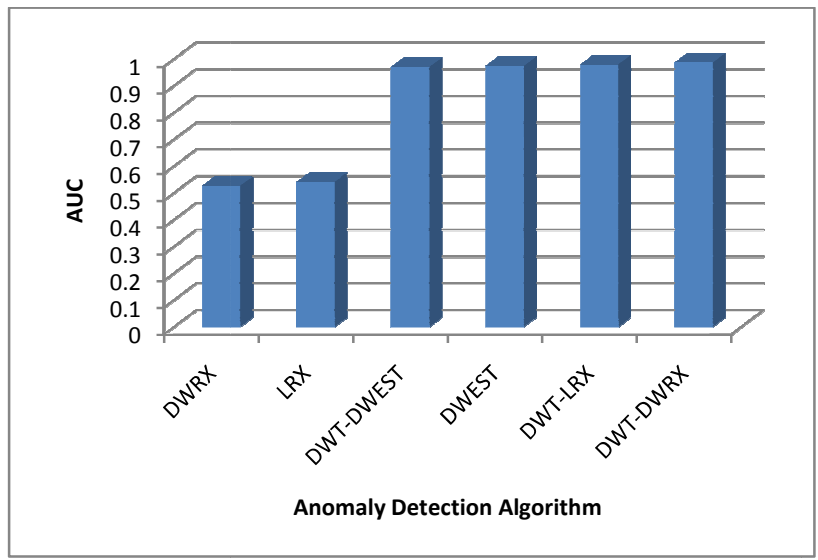

Figure 7. Comparison of AD algorithms applied to Img1

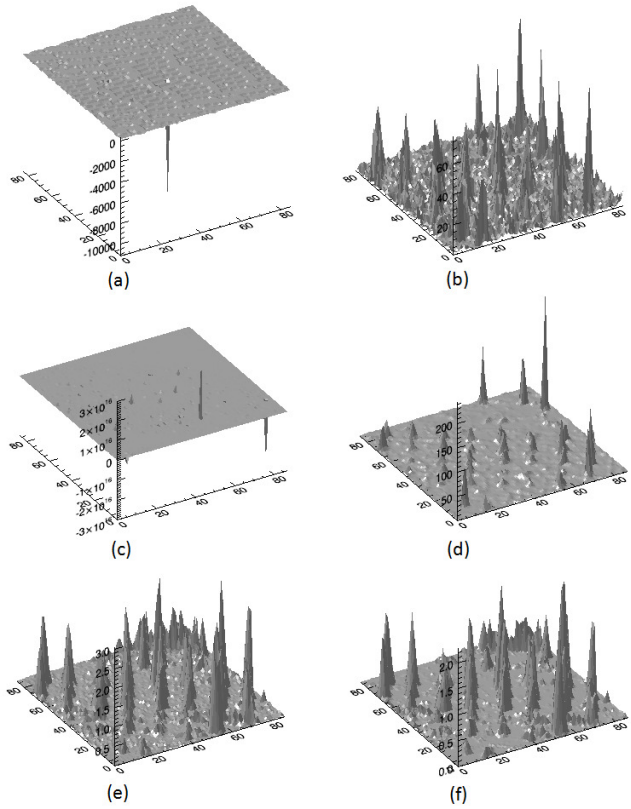

Figure 8. 3D plots of the detection results for Img2: (a) LRX, (b) DWT-LRX, (c) DWRX, (d) DWT-DWRX, (e) DWEST and (f) DWT-DWEST

### 5.3.2. Performance evaluation by Img2

- **Detection Performance**

This sub-image is used to evaluate the performance and runtime of anomaly detectors. Figure 8 shows the detection results of AD algorithms for Img2. Because the ground truth of the targets is not available, the detection performance of anomaly detectors is investigated visually. For this purpose a thresholding step is added at the end of AD procedure. This post-processing step need cut-off threshold, this value can be calculated adaptively using equation (12) [25]:

$$\tau_\alpha = \mu_d + Z_\alpha \times \sigma_d \qquad (12)$$

Where $\tau_\alpha$ is the cut-off threshold value which (at a significant level of α) declares whether a pixel is an anomaly or not, $\mu_d$ and $\sigma_d$ are the mean and standard deviation of anomaly matrix (output of AD algorithm) and $Z_\alpha$ is the *z* statistic at the significant level of α. controlling the number of pixels declared as an anomaly done by $Z_\alpha$. Figure 9 shows the output of thresholding step using the adaptive cut-off threshold which is declared in equation (12). According to Figure 9, the detection performance of RX and DWRX methods is very low and using proposed method improves their performance significantly. In addition, the detection performance of DWEST method does not change significantly.

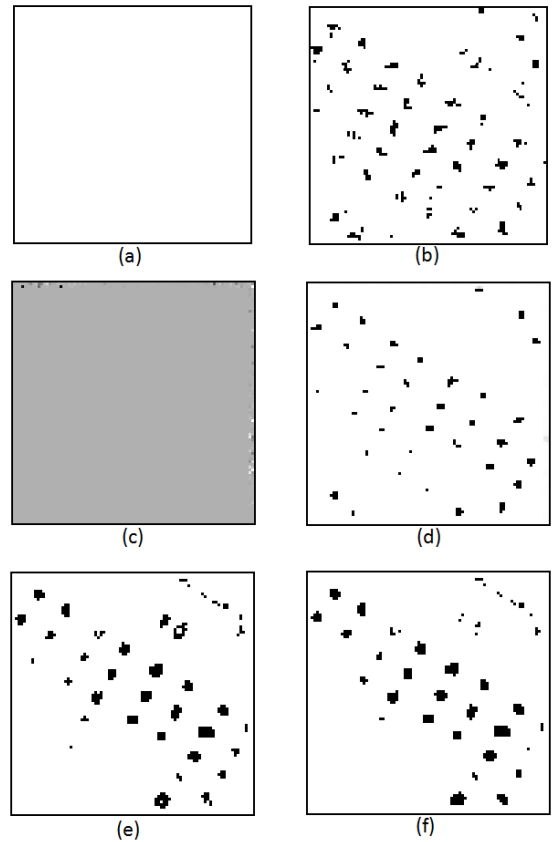

Figure 9. Detection results of methods applied to Img2: (a) LRX, (b) DWT-LRX, (c) DWRX, (d) DWT-DWRX, (e) DWEST and (f) DWT-DWEST

Table 2. Runtime of AD algorithms for Img2.

| AD algorithm | Run Time (seconds) |
|---|---|
| *LRX* | 713.674 |
| *DWT-LRX* | **17.071** |
| *DWRX* | 637.089 |
| *DWT-DWRX* | 19.318 |
| *DWEST* | 1091.173 |
| *DWT-DWEST* | 20.129 |

- **Runtime Performance**

To evaluate speed of AD methods, a computer with "Intel Core i5 2410M, 2.3GHz" processor and 2GByte Random Access Memory (RAM) has been used and runtime of algorithms calculated in equal conditions. Runtime of algorithms is shown in Table 2. The proposed method has an extra time for calculating DWT of the image which equals to 10.8 seconds. This pre-processing runtime is added to the proposed algorithms' runtime. According to Table 2, runtime of AD algorithms with proposed method is much smaller than main methods. The DWT-LRX requires least time while the DWEST is the most time-consuming method.

## 6. CONCLUSION

This work proposed a new method to improve the performance and runtime of current AD algorithms. It uses the Discreet Wavelet Transformation (DWT) as a pre-processing Data Reduction (DR) step. Experimental result on AVIRIS hyperspectral image using ROC curve, AUC and visual investigation showed that this method improved detection performance of the LRX and DWRX methods, significantly. This significant improvement is because of elimination of redundant spectral bands and increasing the separation between anomaly and background signatures using proposed DR method. In addition, proposed method improves runtime of the LRX, DWRX and DWEST methods significantly which is very important in real time application. The conclusion of this study is that the DR pre-processing step can improve the detection performance and speed of anomaly detectors and DWT is an effective DR method for the application of AD in hyperspectral images. Future works include evaluating different wavelet filters and applying the algorithm on other target detection methods in hyperspectral images.